# Nudge for Deliberativeness: How Interface Features Influence Online Discourse

**Sanju Menon**
Department of Communications and New Media
National University of Singapore (NUS)
Singapore
free.feyn@gmail.com

**Weiyu Zhang**
Department of Communications and New Media
National University of Singapore (NUS)
Singapore
weiyu.zhang@nus.edu.sg

**Simon T. Perrault**
Singapore University of Technology and Design (SUTD)
Singapore
perrault.simon@gmail.com

**ABSTRACT**
Cognitive load is a significant challenge to users for being deliberative. Interface design has been used to mitigate this cognitive state. This paper surveys literature on the anchoring effect, partitioning effect and point-of-choice effect, based on which we propose three interface nudges, namely, the word-count anchor, partitioning text fields, and reply choice prompt. We then conducted a 2×2×2 factorial experiment with 80 participants (10 for each condition), testing how these nudges affect deliberativeness. The results showed a significant positive impact of the word-count anchor. There was also a significant positive impact of the partitioning text fields on the word count of response. The reply choice prompt showed a surprisingly negative affect on the quantity of response, hinting at the possibility that the reply choice prompt induces a fear of evaluation, which could in turn dampen the willingness to reply.

**Author Keywords**
Nudges; online discussion; portioning text fields; word count; reply choice prompt; deliberativeness.

**CSS Concepts**
• **Human-centered computing~Empirical studies in collaborative and social computing** • Human-centered computing~Web-based interaction • *Information systems~Collaborative and social computing systems and tools*

**INTRODUCTION**
In recent years, there has been a lot of interest in using interface design to enhance deliberativeness of online discourse (e.g., reviews or discussions) [58][61]. Deliberativeness refers to one's ability to recognize and rebut other's arguments, as well as formalizing one's own.

One significant challenge that prevents users to be deliberative is cognitive load, the mental resources a person has available for completing tasks at a given time. The cognitive load imposed on a person can compromise the quality of arguments produced [42], and selectively interfere with utilitarian moral judgment [17]. The cognitive effort associated with deciding what one wants is particularly high when there is no pre-existing preference for the individual to identify the best option and underlying tradeoffs [52].

The concept of cognitive load has been used extensively within the Computer Supported Collaborative Learning (CSCL) literature, where they use design interventions aimed at reducing cognitive loads, so that students can focus on their primary learning task. Sweller et al. [46] showed that designs can help in the imparting of educational instruction—switching from simultaneous display of information to successive display of information, aimed at reducing loads on working-memory, can lead to increased overall recall of educational materials. Rasmussen and Vicente [34] theorised that human errors related to decision-making can be reduced through system interface design using computational aids that help in improving the functional understanding of a problem domain.

Simple interfaces changes may lead to significant results: Wang et al. [54] found that the degree of "regret" reported during online disclosures (on Facebook) could be reduced, by using behavioral designs targeted at making users more aware of their privacy. Specifically, they used: 1) visual cues, like placing photos of Facebook friends who could potentially view your post at the stage of submission of the post; 2) timers, that allowed a user to cancel a post after submission, within a set period of time; and 3) sentiment feedback, that told users how their post is likely to be viewed, based on a sentiment analysis of its content. They found that the visual cues and timer interventions helped users pause and take stock of their posts. However, they did not find much utility for the sentiment feedback mechanism, as most users reported not using them.

The closest to our attempt comes from Murray et al. [27], who examined the role that design can play in facilitating what they called "social deliberative skills"—a term they use to indicate the capacity to deal productively with heterogeneous goals and values encountered during a collective decision-making scenario. The authors identified five key varieties of social deliberative skills: social perspective taking, meta-dialogue, social inquiry, systems-thinking (complexity thinking), and self-reflection. The aim was to test three stages of reflective tools that can facilitate the abovementioned competencies among citizens. The results of their study showed moderate success at inducing social deliberative skills.

The studies reviewed above all attempt to change user behavior through changing the choice architecture. The design features aiming at the choice architecture, or nudges [51], are to influence choices by manipulating how they are presented, such that certain options from the choice set are more actionable than others. In addition, nudges manipulate the access that individuals have towards choices, so that the ease of use, in relative terms, is slanted towards a particular choice or choices. The nudge paradigm seeks to modify the choice architecture, so that we can effectively intervene and moderate the influence that the automatic mode of thinking has on our decision process. A key feature that makes nudges different from other cognitive design mechanisms is that nudges preserve freedom and complexity. Thaler and Sunstein [51] include 3 main requisites for a nudge: 1) it can be used without incurring much cost to the user; 2) it is easy to opt out of a nudge (or in other words, it is easy to turn a nudge off); and 3) it does not distort existing incentive structures in any significant way. Therefore, fines and penalties do not constitute a nudge, because they change the incentivisation structure. In next section, we turn to interface design literatures to discuss three types of nudges, namely, anchoring, partitioning, and reply choice, and their potential to nudge for deliberativeness.

The contribution of this work is two-folded:

- The results of a user study investigating how three simple interface nudges (word count anchor, partitioning text fields and reply choice prompt) impact the deliberativeness of online discourse
- A set of design guidelines on how to use these interface nudges in similar contexts and other types of online discussions.

## BACKGROUND AND RELATED WORK

Nudging has been extensively studied in both HCI and other fields such as public administration and communication. We first start with a brief summary of nudging in HCI and the different contexts in which nudging has been used, followed by a discussion on the three main aspects of our focus: Anchoring, partitioning and point of choice.

### Nudging in HCI

Previous work in HCI related venues (e.g. CHI, CSCW) explored the different ways to nudge in various contexts. Caraban et al. [7] presented a review of 71 HCI works in nudging, which allowed them to include 23 distinct mechanisms of nudging that they grouped into 6 categories.

The works considered in that review greatly vary in terms of their level of reflectiveness and transparency. For instance, reminding users of the consequences is meant to encourage users to be more reflective – Harbach et al. [18] changed the permission dialogue of Google Play in order to illustrate and remind the user of the consequences of their choices for the sake of privacy. Another example is to provide multiple viewpoints to mitigate the confirmation bias. Park et al. [30] designed NewsCube, an application that collects different points of views for an event and offer an unbiased clustered overview to the users.

In our context, the goal is to improve the overall deliberativeness of online discourse, when users are discussing a controversial issue [31]. For this specific case, we hope that different mechanisms potentially improve both the quality of the posts (in terms of how they take into account the other party's arguments and how they form their own) and the quantity of text in the posts (though higher number of words does not always correlate with higher quality). To the best of our knowledge, the factors we considered were not examined in similar contexts by previous works.

### Deliberativeness

Deliberativeness is a core value of public deliberation and focuses more on how ideas are exchanged and discussed, and thus "involves recognizing, incorporating, and rebutting the arguments of others... as well as justifying one's own". [12]. It is a composite measure, which encompasses has been defined in diverse ways in the literature [32],[44][45]

Trénel [49] proposed a coding scheme for deliberativeness with 8 dimensions. For this work, we decided to focus on two dimensions: (1) *Respect for others' arguments* and (2) Constructiveness, as the *number of arguments expressed*. In addition, we also measured the number of words per response, which while not a key part of deliberativeness, is an overall interesting indicator of the evolution of a conversation on online discussions.

### Anchoring

Anchors are understood as default options that have the effect of influencing an individual's decision, typically towards the direction of the default. In this version, an anchor can be viewed to be exploiting what is called the status quo bias [21]. Work on the phenomenon of cognitive load has shown that anchoring matters most when the stakes are small, or when people do not need to spend too much effort, because the importance of making the decision is viewed as low [26].

One explanation of why defaults are chosen is based on the 'availability heuristic'. Here people decide on an estimate for the frequency of an event, or the likelihood of it occurring, based on the ease with which instances or associations regarding said events are recalled in the mind. Often, default options are implicit reflections of some external reality; wherein the default option is appropriate for a wide array of decision-makers. An example of this is "Save More Tomorrow" from Thaler and Sunstein [51] (2008), where they set default options related to health care plans, with the aim of helping citizens make safer choices regarding their health care. Another somewhat extreme example of this is from Redelmeier and Shafir [38], who found that adding a new option increased the probability of choosing a previously available alternative, suggesting that default anchors can work across successive choice scenarios.

Dinner et al. [9] argue that defaults work on the endowment effect, whereby the decision-maker may act as if they have already chosen the default option, and consider it as a reference point. Park et al. [29] showed that the endowment effect also imbues a default choice with more value— the value for an object decreases when it is owned [50]. Roumbanis [41] looked at an application of the anchoring effect within deliberating panels in Swedish Research Council panel groups. This application targeted the anchoring phenomenon in relation to expert knowledge, and how it affects academic negotiations during consensus-building talks. Englich and Mussweiler [10] have looked at how the anchoring effect can bias judgments on legal questions, by both layperson and judges during jury trials.

For this study, we focus on the application of anchors in the context of task-motivation situations, wherein anchors influence the time and effort individuals choose to invest, towards meeting the goals of a task [47] (p. 213). Other research suggests that anchors would change the task performer's frame of reference regarding what would be an appropriate goal. Most task performers acting alone use their most recent performance on the task as the basis for their goal [19], and with the lack of incentives, they do not push themselves toward higher levels of performance. By introducing an anchor, the task performer's frame of reference may shift from the previous performance toward the value introduced by the anchor. A judgmental anchor may also change individuals' beliefs about their capability to perform well on the task. Cervone and Peake [8] demonstrated that arbitrary anchor values (i.e., numbers picked at random) changed participants' self-efficacy expectations regarding their performance on problem-solving tasks. Consequently, we might expect goal-based anchors to influence the self-efficacy that individuals hold toward a task. Self-efficacy, in turn, has been related to reflective attention [2] (p. 123) and comprehensive evaluation, or the generation of counterfactuals. Work in psychology has shown a link between self-appraisals of efficacy and empathy for others [3] and perceived autonomy [43]. Hence, goal-based anchors could potentially serve as a nudging mechanism for deliberativeness..

We thus decide to examine the word count anchor, an anchor located below a text field that contains a progress bar and a word count to show the length of a post to the user. Our first hypothesis is as follows:

H1: Word count anchor increases deliberativeness.

**Partitioning**

The field of cognitive psychology has also looked into how humans reason under uncertainty, more specifically, the type of biases individuals exhibit when they are forced to make a decision with limited or no information. One such decision scenario is when individuals must decide how to allocate resources among options that they are uncertain about. One important facet of such a choice architecture is the way in which the set of options often exists in a partitioned form—in groups or categories. This feature of a choice environment can have a significant implication for the choice behaviour. Grouping options can assist consumers, in part because categories provide important information about the shared attributes of items in that category [20]. Likewise, presenting choices in narrower or broader brackets may determine whether consequences are considered in isolation or in combination [37]. When people have limited cognitive resources to allocate to a choice task, they typically prefer to allocate their cognitive resources evenly to each group or category that has been identified. This phenomenon is called partitioning.

Prior research in diverse domains has shown that partitioning creates vivid categories that can influence allocations involving simultaneous choices. Partitioning is thought to arise from a desire to hedge against risky or uncertain prospects, and minimize the potential for post-choice regret [4],[35],[36]. Thaler and Sunstein [51] report that employees asked to allocate retirement investments prefer to do it evenly over various categorical options, such as stocks, bonds, and real estate, when these separate categories are made identifiable. Wansink et al. [55] showed that partitioning of online order forms could alter the mix of products a person chooses to purchase. Tannenbaum et al. [48] found that partitioning of menu items affects the way individuals spend "attentional" resources towards them. They also argue that partitioning can end up conveying information about the relative popularity or suitability of different groups or categories in the partition. For instance, menu partitions could signal information about descriptive norms, with greater granularity or sub-partitions indicating the greater popularity of a group. In recent years, the partitioning of web interfaces has been used to influence consumer choice [14]. They found that partitioning was a robust effect working across a wide array of situations—allocation of money, chance gambles and investments. They reasoned such diversification across different partitions to be motivated by a desire to hedge against uncertainty in decision-making. Partitioning effects have also been observed in cases where people are required to allocate different degrees of belief among possible events that might occur. For instance, when assigning probability estimates to unfamiliar events, people tend to invoke the "principle of insufficient reason" [57], treating all possible events as equally likely.

The pervasive tendency of partitioning towards even allocation can be used to design for deliberativeness. By splitting one input text field into more input fields, a compulsion for even allotment to multiple aspects will drive users to offer more perspectives, in an attempt to fill in each text field. The weight that different attributes receive depends on how they are partitioned: attributes that are displayed as separate categories tend to receive greater weight, whereas those that are grouped together under umbrella categories are discounted as less important [25]. Using this idea of partitioning, we expect that splitting the input text field can increase the salience of different types of content that can go into a response. For example, splitting a broad reply field into separate fields like 'position' (what your position on the argument is, for example 'agree' or 'disagree'), 'explanation' (why you have taken that position) and 'evidence' (why evidence can back up your position). We thus propose our second hypothesis.

H2: Partitioning text fields increases deliberativeness.

**Point of Choice**
Point of choice prompts are a possible tool for behaviour change [16]. Point-of-choice prompts function by interrupting habitual behaviour at the point of its occurrence, allowing for the substitution of habitual behaviour with a more desired alternative [56]. Point of choice prompts are usually very transient and do not hold the attention of the choice-maker for too long. It is therefore less capable of motivating new types of choices; instead, it is more likely to work by reminding individuals of their prior intentions to be more active at a time and place where they can fulfil them, helping translate intentions into actual behaviour [33]. According to the cognitive load theory [46], prompts should not produce additional extrinsic cognitive load. According to Berthold et al. [5], point of choice prompts can be helpful in supporting learning and the use of more deliberate cognitive processes. They found that taking temporally-presenting prompts closer to the time of learning was more effective in facilitating learning outcomes, than presenting the same prompts temporally apart, which requires holding them in working memory until needed. Puig-Ribera and Eves [33] showed that a prompt positioned at the point of choice between the stairs and the escalator can encourage pedestrians think more consciously about their health, and with regard to taking the stairs. According to them, "[by] changing the contextual cues, point-of-choice prompts provoke deliberation by pedestrians about the behaviour rather than choosing the escalator in a 'mindless' manner'". Evans et al. [11] showed that point of choice prompts can be used to reduce the amount of time spent sitting at work.

The social presence literature has shown that being aware of the presence of others can prime individuals to be more empathetic [6]. Research on digital gaming has also shown that priming for social presence can increase empathy for others you play with. Point of choice prompts aimed at priming the social presence of others may have the effect of eliciting increased other-regarding orientation from users. We thus propose the use of a reply choice-prompt, which asks users if they would like to have the author of the governing argument (a fictitious one) reply to their response. This may have the effect of priming other-regarding orientation, before individuals start typing their responses.

H3: Reply choice prompt increases deliberativeness.

**EXPERIMENT**
To investigate the potential effect of Point of Choice, Partitioning Text fields and Word Count Anchor, we ran a controlled experiment in Singapore.

**Method**
The study was designed as a between-subjects experiment, using a 2×2×2 factorial design with three binary independent variables: Reply choice prompt *{ present, absent }*, Partitioning text fields *{ present, absent }* and Word count anchor *{ present, absent }*. There were 8 experimental conditions in total as seen in Table 1.

| Condition | Reply choice prompt: (asking participants whether they want the poster to respond to them) | Partitioning text fields: (input field split into 3 subfields) | Word count anchor: (real-time progress bar showing word counts, capped at 400 words) |
|---|---|---|---|
| 1 | ✗ | ✗ | ✗ |
| 2 | ✓ | ✗ | ✗ |
| 3 | ✗ | ✓ | ✗ |
| 4 | ✗ | ✗ | ✓ |
| 5 | ✓ | ✓ | ✗ |
| 6 | ✗ | ✓ | ✓ |
| 7 | ✓ | ✗ | ✓ |
| 8 | ✓ | ✓ | ✓ |

**Table 1. Summary of 8 experimental conditions.**

**Power analysis**
Prior to the recruitment of participants, we conducted a power analysis to ascertain the threshold sample size for an 8-group ANOVA study with effect size 0.5 and power 0.95, where the threshold was found to be 55.

**Participants**
We recruited 80 participants for this study. The 80 participants (62 females and 18 males) aged 21-31 (M=22)

were recruited in two rounds. In the first round of recruitment, they were incentivised to participate with 10 SGD cash or voucher rewards. The second round of recruitment involved a lottery system, where the winner would get a cash award of the equivalent of 30 USD. They were recruited through snowball sampling. Initial recruitment emails were sent to college students attending two large introductory classes. Participants were predominantly of Chinese (46 out of 80) or Malay (22 out of 80) ethnicity. Seven participants identified as Indian and 5 as 'others'. 92.5% of the participants reported themselves as currently doing their undergraduate studies, while 2.5% reported themselves as graduate students. In terms of employment, all reported themselves as either students or with no work experience.

**Task and Materials**

The interface is designed to handle both graphical and text-based input interactions. A total of 8 different interface prototypes were tested for this study (Table 1). Each design captured a different type of choice architecture. The tools or nudges used to manipulate the choice architecture were: Partitioning Text Field, Reply Choice Prompts, and Word Count Anchor. During each condition, a different design intervention was used to manipulate the choice architecture of content contribution. But participants are all supposed to reply to a previous post. The post's content was fixed for every participant and read as follows:

*"There has been an increase in the foreign and migrant population in Singapore in recent years. This may be due to the fact that companies are choosing to hire immigrants over Singaporeans citizens. The influx of foreigners has also likely resulted in a situation where public services like SMRT are being increasingly strained and overwhelmed over time. Immigration has also likely created problems with cultural integration, possibly leading to a loss of identity among Singaporeans. It may be desirable that the government provide monetary incentives to companies so that they are encouraged to hire more Singaporeans & thereby facilitate better social-integration with immigrant populations."*

The screenshots for each of the individual conditions are presented below. Firstly, the study used a control interface (see Figure 1) to test the effects of using nudges, as compared to when there were no nudges. The control interface used in the study is a simple input panel for each of the three types of deliberative content. This interface does not contain any of the design interventions aimed at manipulating the choice architecture of the user.

The reply choice prompt (Figure 2) interface had a prompt at the top of the input text field, so that the nudge was easy to identify to the user.

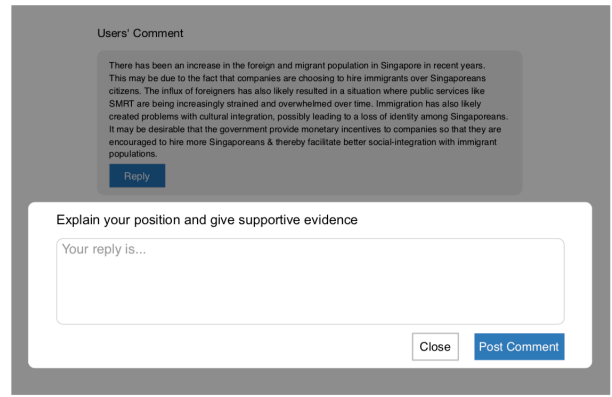

**Figure 1. Control Interface (condition 1), with a simple text field and two buttons (close and post).**

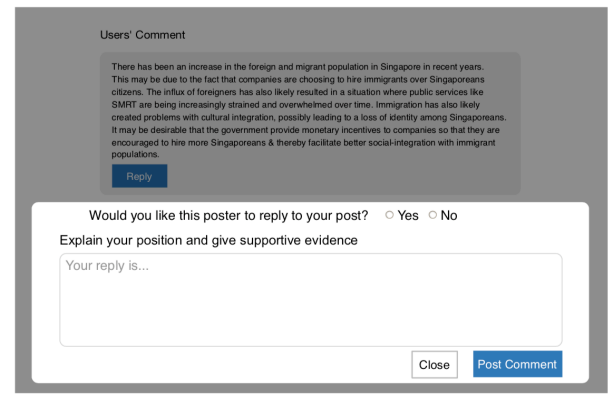

**Figure 2. Interface design with the reply choice prompt (condition 2) at the top of the reply pop-up.**

The partitioning text fields interface had three separate sentence openers: the first one had "Your position is" as a sentence opener (a ghost text, which disappears when the user clicks inside the text field). The second partition had "Explain your position here", and the third partition had "Supportive evidence" as sentence openers, respectively. The idea was to give individuals three separate types of points they could raise, possibly driving them to allocate an equal number of time. An important feature to note here is that the Partitioning text fields (Figure 3) always included ghost-text (position, evidence and explanation). For the control, the title "Explain your position and give supporting evidence" was added in the control interface (Figure 1), so that any effect related to the presence of these textual instructions or clues did not affect the results.

The word-count anchor interface (Figure 4) had a progress bar-type interface, which counted the number of words typed in, with a maximum of 400 words. However, this did not mean that they could not write more than 400 words, as it was only the limit of the progress bar feature; there was no limit to how much they could write. For far comparison, the word-count anchor maxed out at 400 words even in the Partitioning text field case, where there were 3 text fields for input. The word count was calculated as simply the sum of the words in each text field.

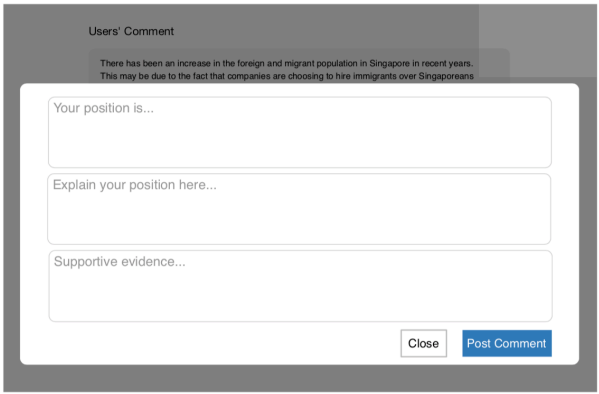

**Figure 3. Interface design with the partitioning text fields (condition 3).**

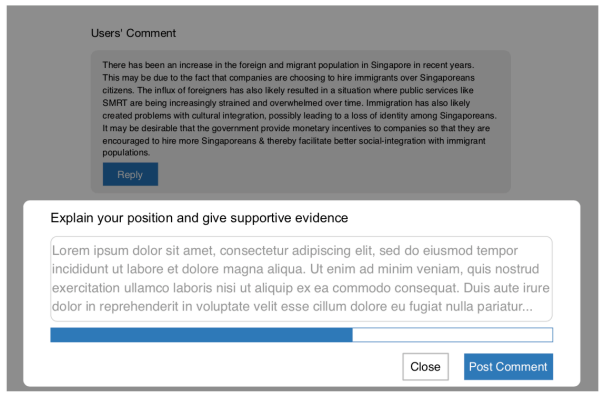

**Figure 4. Interface design with the word count anchor only (condition 4). The anchor is shown at the bottom, as a blue progress bar.**

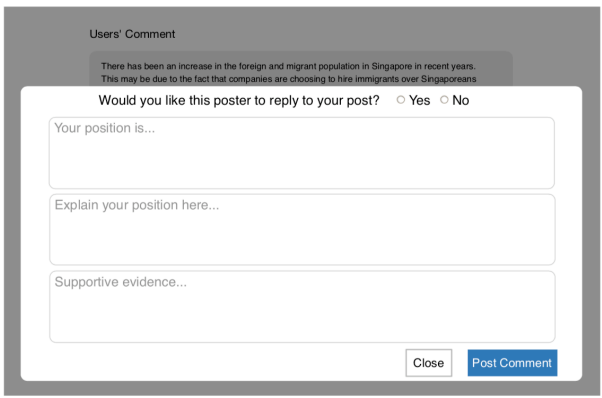

**Figure 5. Interface design with the reply choice prompt and partitioning text fields (condition 5).**

In conditions where two or more nudges were simultaneously presented (Figure 5, Figure 6, Figure 7, Figure 8), the nudges appeared in the same location as in the individual cases.

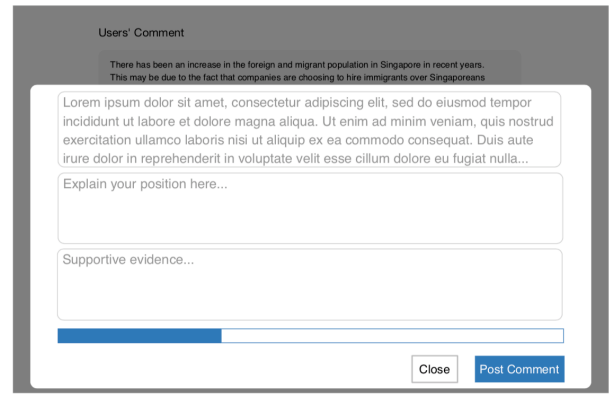

**Figure 6. Interface design with the partitioning text fields and word count anchor (condition 6).**

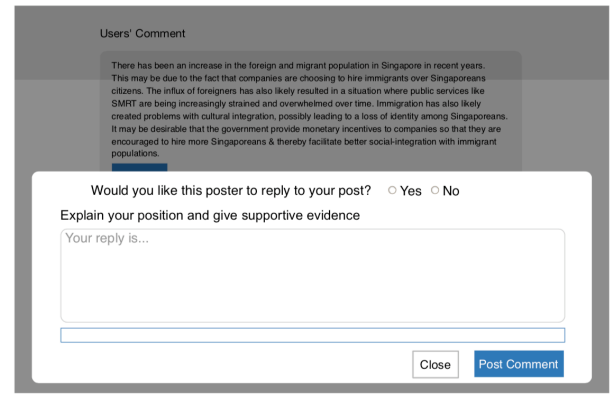

**Figure 7. Interface design with the reply choice prompt and word count anchor (condition 7).**

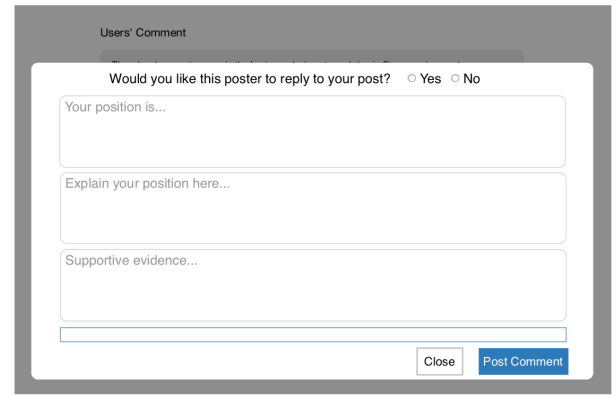

**Figure 8. Interface design with all three features: the reply choice prompt, partitioning text fields and word count anchor (condition 8).**

### Procedure

Participants were recruited via email invitations and through snowball sampling, wherein the purpose of the study was described as an interaction design study. No information about the experimental conditions was provided to the participants. Each participant was asked to respond to a post on the topic of immigration, but there were no explicit instructions given on how to interact with the post on different interfaces. For instance, there was no mention of the word count anchor feature, or how to make a selection on the reply choice prompt, or what the partitioning text fields were for. After they furnished their responses to the post, they were required to provide arguments for and against the current government policy on immigration.

**Dependent variables**

Deliberativeness is a multi-dimensional concept [59]. We decided to operationalize it through three measures (see Background and Related Work):

1. Response word count,
2. Respect for arguments,
3. Argument repertoire.

Response word count is a simple auto count of words typed into the reply areas. The rest of measures were derived from a content analysis of the replies. Two coders carried out the coding. The primary coder was a PhD student who majored in communication. The secondary coder was a working professional who had been living in Singapore for over 3 years and was familiar with the immigration topic. Cohen's Kappa was run to determine if there was agreement between the two coders' judgment: Respect for argument (Kappa = 0.910, p <0.001); Argument repertoire - Against Immigration (Kappa = 0.791, $p<.001$); Argument Repertoire - For Immigration (Kappa = 0.840, $p <.001$).

Participants' respect for argument was coded at three levels: 1) ignores the argument present in the governing argument—this indicates that the participant did not engage with the thoughts expressed in the governing argument; 2) includes argument but not treated with seriousness—this indicates that the governing argument was mentioned in the response but only in passing, without offering any critique; and 3) includes argument and talks about how it is right or wrong: This indicates that the response engaged with the governing argument seriously, by critiquing the points raised in it.

Argument repertoire was coded by counting the number of non-redundant arguments regarding each position (either for or against immigration). The ideas produced along the two positions were combined. Redundant arguments that were repeated across positions were counted only once.

**Manipulation Check**

For studies that involve interface nudges, it is important to test if the implemented features were being identified as such. The intention here is to understand if the manipulations of the interface features were indeed noticed by users. The check was similar to the method used by Wang et al. [53]. The nudges were first evaluated internally by the authors and a university student, to identify possible problems with the design. For instance, in the first version of the design for the reply choice prompt, the question "Would you like this poster to reply to this post" appeared at the bottom. From our review, we realised that this probably would not have had the intended effect, as participants may have only noticed this prompt after they had started writing in the text fields. Therefore, it was decided to change the reply prompt to appear at the top, so that it would be noticed before the participant started writing their response.

In line with Wang et al. [53], we then proceeded to test if the nudges and their respective use-cases were being identified by the users. A group of 5 participants were recruited and each was separately shown the manipulated nudge designs (8 in total) in succession. They were given 1 minute of time to use the interface, and quizzed about what they observed in relation to the interface. All participants could identify the word count anchor each time and were able to associate it with the quantity of words typed in the text field, and also understood that the feature was not a progress bar to complete before sending their post. Reply choice prompts were also easily identified each time and were accurately associated with the choice to have the poster reply to them. Partitioning text fields were also identified each time and were associated with giving multiple replies. However, one participant opined that the partitioning text fields was a little confusing to use, as they were not sure what counted as evidence (the third category in the partition). It is important to note that the manipulation check here is only related to whether the nudges were identifiable or noticeable in terms of their functionality. A check done on whether the manipulations were of a sufficient degree to be able to affect the cognitive markers was not warranted, as this should not be the focus of manipulation checks [28]. The manipulation checks were mostly aimed at ascertaining whether the intervention does indeed qualify as 'transparent'.

**Results**

Given our design with three independent variables and number of dependent variables, we will summarize the results and focus the analysis on the significant ones.

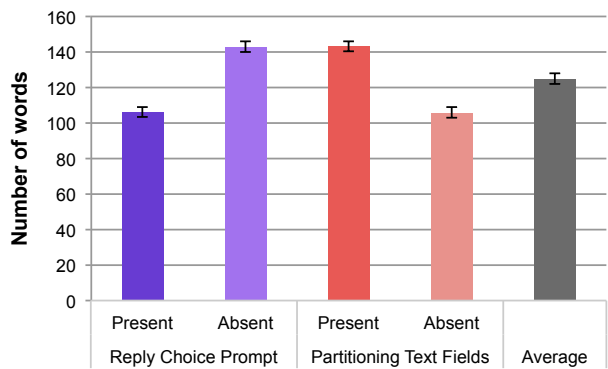

**Figure 9. Average response word count for Reply Choice Prompt and Partitioning Text Fields conditions. Error bars show 95% confidence intervals.**

*Response word count*
A three-way ANOVA was conducted to examine the effects of our three independent variables on response word count. The average response contained 124.6 words.

There was a statistically significant main effect of *Partitioning Text Field* ($F_{1,72} = 8.00$, $p <.05$, $\eta^2= 0.1$), such that the average response word count was higher when *Partitioning Text Field* was present ($M=143.40$), versus when it was absent ($M=105.75$).

There was a statistically significant main effect of *Reply Choice Prompt* ($F_{1, 72}=7.52$, $p <.05$, $\eta^2= 0.01$), such that the average response word count was lower when Reply Choice Prompt was present ($M=106.32$ words), versus when it was absent ($M=142.82$ words).

We did not find any statistically significant main effects of *Word Count Anchor* ($p >.05$) or any interactions. The results are summarized in Figure 9.

*Respect for arguments*
An ordinal logistic regression analysis was conducted to predict respect for arguments using *Reply Choice Prompt, Partitioning Text Field* and *Word Count Anchor* as predictors. A test of the full model against a constant-only model was statistically significant ($p<.05$), indicating that the predictors, as a set, reliably distinguished between levels of respect for argument.

The Wald criterion demonstrated that *Word Count Anchor* had a close to significant main effect on prediction ($p= .082$). The exp(B) value indicated that, when the *Word Count Anchor* was absent, the odds ratio of showing respect for other arguments was 0.184. In other words, when a word count anchor is present, it is 5.6 times more likely to score better in respect for argument. No other significant main and interaction effects were found.

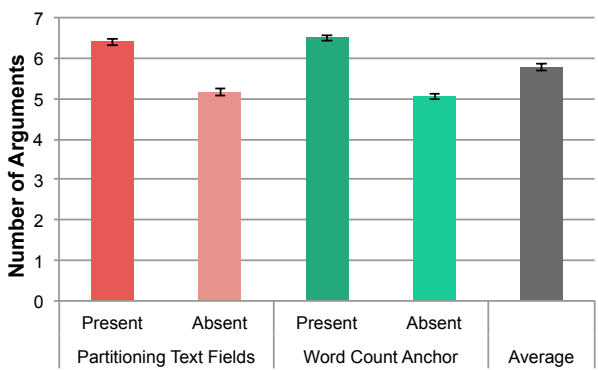

**Figure 10. Average number of arguments per response for each condition. Error bars show 95% confidence intervals.**

*Argument repertoire*
A three-way ANOVA was conducted to examine the effect of our independent variables on argument repertoire. On average, participants provided a total of 5.77 arguments.

There was no statistically significant main effect of *Reply Choice Prompt* ($p>.05$).

There was a statistically significant main effect of *Word Count Anchor* ($F_{1,72}=15.01$, $p<.05$), such that the average number of arguments in the argument repertoire was higher when the word count anchor was present ($M= 6.50$), versus when it was absent ($M= 5.05$). The results are summarized in Figure 10.

There was also a statistically significant main effect of *Partitioning Text Field* ($F_{1,72}=11.16$, $p<.05$), such that the average number of arguments in the argument repertoire was higher when *Partitioning* was present ($M=6.40$), versus when it was absent ($M=5.15$).

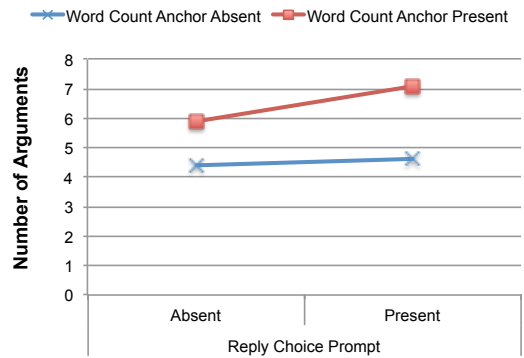

**Figure 11. Average number of arguments depending on the presence of both Reply Choice Prompt and Word Count Anchor.**

There was a statistically significant interaction between *Reply Choice Prompt* and *Word Count Anchor* ($F_{1,72}= 7.14$, $p<.05$, $\eta^2= 0.09$). A plot of marginal means (see Figure 11) shows that although word count anchor in general increases argument repertoire, this increase is conditioned by *Reply Choice Prompt*. In other words, the increase is smaller (from 4.40 to 5.90) when *Reply Choice Prompt* is absent; the increase is larger (from 4.60 to 7.10) when *Reply Choice Prompt* is present (Figure 11).

There was also a statistically significant interaction between *Partitioning Text Fields* and *Word Count Anchor* ($F_{1,72}=7.85$, $p<.05$, $\eta^2= 0.09$). A plot of marginal means (see Figure 12) shows a similar pattern as the one in last paragraph. Although word count anchor in general increases argument repertoire, this increase is conditioned by *Partitioning Text Field*. In other words, the increase is smaller (from 4.90 to 5.20) when *Partitioning Text Fields* is absent; the increase is larger (from 5.10 to 7.80) when *Partitioning Text Fields* is present. No other significant interaction effects were found.

In summary, **H1 is partially supported. H2 is partially supported. H3 is rejected**.

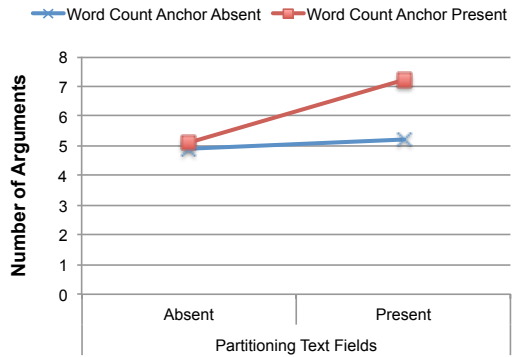

**Figure 12. Average number of arguments depending on the presence of both Partitioning Text Fields and Word Count Anchor.**

## DISCUSSION

In this section, we discuss our results in light of previous works, but also in terms of implications for studies on increasing deliberativeness of various online discourses.

### Word count anchor

Our findings on word count anchor confirm some previous results, but also go against some other findings.

*Confirmed results*

First, our results showed a significant effect of the word count anchor on increasing number of arguments, as well as respect for arguments. This supports Switzer and Sniezek's [47] (p. 213) work, which showed a positive connection between goal-related anchors and effort. This also supports Bandura's [2] (p. 123) findings, which suggest that anchoring can affect reflective attention. The finding also lends credence to Cervone and Peake's [8] argument, suggesting a link between the generation of counterfactuals and anchoring.

*Differences from previous work*

This, however, goes against other findings. Manosevitch [24] for instance, showed that when reflective cues (like a statement suggesting the importance of thinking about issues) were used within news media articles, they failed to show any significant effect on argument repertoire. This suggests that, perhaps the best way to prime people to think of more arguments may not be to do it directly, but rather by using other features as a prime. Rietzschel et al. [40] showed that priming (in the form of probe questions) could help increase the quantity of ideas produced during brainstorming sessions. They argue that priming can help by increasing the accessibility that participants have to their own knowledge (stored in long term memory). The increased availability of knowledge from the prime could have opened accessibility up to more arguments, and hence make better their ability to reason. This may have been the reason for the increase in the respect for arguments score when the word count anchor is used in our study.

*(Lack of) Impact on response word count*

However, the word count anchor did not increase the response word count. This suggests that having more 'argument ideas' does not necessarily translate to a willingness to write more words. It also shows that the word count anchor did not work in a like-breeds-like fashion. This has interesting consequences: for instance, it points to the possibility that people write based on relevance or on what they think is an appropriate response to the argument, and not based on what ideas they have. In a comparable study, Locke and Latham [23] reported that the type of anchoring could change the nature of performance on a task. The task in consideration was 'proofreading', where results showed that explicit anchors have a higher effect on task speed, and task accuracy. The explicit anchors they used were instructions like "do your best", whereas implicit anchoring was achieved by asking individuals to construct a four-word sentence from a randomly-positioned word list—this list in turn contained words related to achievement like strive, accomplished, etc. The present study followed an implicit goal-setting anchor, as participants were not instructed to write as much as they could, or until the word count anchor bar was filled up. Hence, the word count anchor probably only helped as an implicit anchor and was not able to encourage higher quantity of response.

### Partitioning text fields

The partitioning text fields showed a significant positive effect on the word count of the response. This effect could be because the feature was able to nudge participants to write an equal amount of words in each section. The partitioning text fields triples the writing area (in terms of visible area), thereby possibly prompting more writing on the part of the participant. Partitioning is supposed to work through 'naive diversification' [21], which in turn works to create a more even spread across each of the categories in the partition, like the evidence and explanation categories. The partitioning text fields did not have a significant main effect on other dependent measures, though. This may have been due to an increase in cognitive load, introduced by the increased salience associated with each text field category (claim, explanation and evidence). It is possible that the increased salience increased the overall complexity of the task, reducing the amount of available working memory needed. Lavie and Fockert [22] have shown that low-priority but highly salient cues can act as a distraction on tasks that require more working memory. The partitioning text fields could have presented salient cues with regard to the categories of response, while simultaneously distracting the subject from arguing actively during their response.

### Reply choice prompt

Reply choice prompt showed a main effect of reducing response word count, contrary to our expectation. It is possible that the reply choice prompt introduces an element of social judgment, because it asks the individual if they would indeed like the writer of the first post to respond to them (all participants ticked "yes"—indicating they wanted a response). Since the participants were selected through snowball sampling, they probably felt compelled to say yes, as the writer might be someone they knew. Nevertheless, social judgment has been shown to decrease the ability of individuals to be more expressive and creative [1].

The lesson here is that, while the reply choice prompt was meant to induce a sense of commitment on the part of the participant (to get them to treat the interaction as a commitment to a longer conversation, and hence, be driven to be more engaging), it may have also made them feel more awkward about raising their opinions freely. Thus, the introduction of the reply choice prompt seems to have introduced new complications to the choice architecture than what we had hoped for.

### Dependence between factors
In addition, both partitioning text fields and reply choice prompt showed a significant integration effect with word count anchor on influencing argument repertoire. The two patterns are consistent: when word count anchor is combined with either partitioning text fields or reply choice prompt, the number of arguments increases much more than when word count anchor works alone by itself.

## IMPLICATIONS FOR OTHER CONTEXTS
As a follow-up from the previous section, we discuss how our results may be used in different contexts. Our task was very specific (responding to a controversial topic), and it is unclear how our results could generalize in other contexts. The goal of this section is to suggest areas for future work that could potentially leverage our results. As such, more research would be needed to validate these claims.

### Better datasets on Crowdsourcing platforms
In a crowdsourcing scenario where the workers' answers are used as a training data set for a machine learning/deep learning system, the use of partitioning text fields might increase the size of each sample (response) and the number of arguments used in the text. Adding word count anchors could also lead to a higher number of arguments. As such, condition 6 could be the best choice.

### Higher quality reviews on Help platforms
On help platforms or platforms that require genuine and informative reviews (e.g., PCS, restaurant review websites/applications), the response length is not the main focus. Instead, receiving a well-explained and potentially argumentative response is better than a long, vague response without clear explanations. In that context, we would suggest to use a design combining all three factors: partitioning Text Fields not only on its own would lead to more argumentative responses, but also could positively interact with Word Count Anchor to increase argument repertoire. Reply Choice Prompt might also interact with Word Count Anchors and increase the quality of the responses. The design used in Condition 8 (Figure 8) would thus be a good choice.

## LIMITATIONS
One of the limitations of this study could be the prevalence of majority student participants. Since the interface features were relatively simple to use, only a basic level of internet skills was expected from the participants. Although undergraduates are expected to have higher level ICT skills compared to the general population [15], there is no reason to think that the general population would have trouble using the interface. Another potential limitation is the higher percentage of female participants (62 out of 80). Despite this, we found no literature that has systematically studied gender effects on the efficacy of nudge-type interventions. Reisch and Sunstein [39] reported that females (polled in 6 European countries) have a slightly more positive valence toward nudge interventions. However, it is not clear if this difference could have any impact on the responses given by our participants. Another limitation is the unique Singapore context [60], which could influence how people argue in response to the immigrant issue. While we do believe that our results could be generalized to other English-speaking countries, more work would need to be done to confirm this.

## CONCLUSION AND FUTURE WORK
The overall implication of this study is that interface nudges have a moderate level of success, in terms of enhancing deliberativeness of online discourse. Other than the individual effects of each nudge, our study suggests that combining multiple nudges may have better effects, assuming that there are not too many of them that may overwhelm users' cognitive load. Design-wise, depending on the needs of designers to increase the quantity or quality of user input, different nudges could be used. Future work should invent and examine more nudges that could facilitate users' sympathy or respect towards other people and different viewpoints.

## ACKNOWLEDGMENTS

This research is based on the first author's PhD dissertation, which is partially supported by Singapore's Ministry of Education (MOE2013-T2-1-105). Views expressed are those of the authors alone and do not necessarily reflect opinions of the sponsoring agencies. The authors want to extend sincere appreciation to all the team members who have worked on this project. Also many thanks to the conference committee members (ACs) for recognizing the value of this work.